\documentclass[%
 aip,
 amsmath,amssymb,
 reprint,
]{revtex4-1}

\usepackage{graphicx}
\usepackage{dcolumn}
\usepackage{bm}

\usepackage[utf8]{inputenc}
\usepackage[T1]{fontenc}
\usepackage{mathptmx}
\usepackage{etoolbox}

\makeatletter
\def\@email#1#2{%
 \endgroup
 \patchcmd{\titleblock@produce}
  {\frontmatter@RRAPformat}
  {\frontmatter@RRAPformat{\produce@RRAP{*#1\href{mailto:#2}{#2}}}\frontmatter@RRAPformat}
  {}{}
}%
\makeatother
\begin{document}

\preprint{AIP/123-QED}

\title[Reduction of magnetic-field-induced shift in quantum frequency standards based on CPT]{Reduction of Magnetic-Field-Induced Shift in Quantum Frequency Standards Based on Coherent Population Trapping}

\author{V. I. Vishnyakov}

\author{D. V. Brazhnikov}%
\email{x-kvant@mail.ru}

\author{M. N. Skvortsov}

\affiliation{ 
Institute of Laser Physics SB RAS,
15B Lavrentyev Avenue, Novosibirsk 630090, Russia
}

\date{\today}

\begin{abstract}
We investigate the magnetic-field-induced frequency shift (MFS) of the clock ``0--0'' transition in the microwave quantum frequency standard (atomic clock) based on coherent population trapping (CPT) in $^{87}$Rb vapor. To scan the CPT resonance and to form the error signal, a method analogous to the Pound-Drever-Hall (PDH) technique in the optical frequency range is employed, where the modulating frequency ($f_m$) significantly exceeds the resonance linewidth (FWHM). The experiments demonstrate that this technique offers brilliant capabilities for controlling the sensitivity of the clock transition frequency to magnetic field variations in the vapor cell compared to the conventional method with low-frequency modulation ($f_m$$\,\ll\,$FWHM).  Specifically, the PDH technique provides several optimal values of the bias magnetic field generated by the solenoid, at which the ``0--0'' transition frequency exhibits extremely low sensitivity to small variations in the external magnetic field. Furthermore, these magnetic field values can be easily adjusted by changing $f_m$, which is relevant for optimization of the atomic clock's operating regime. The experimental results show that by using the PDH technique, the influence of MFS on the clock transition can be suppressed down to $\approx\,$$3.2$$\,\times\,$$10^{-13}$$\delta B^2$~mG$^{-2}$. These findings can be leveraged both to relax stringent requirements for magnetic field shielding in state-of-the-art CPT-based miniature atomic clocks (MACs) and to build a new generation of such clocks with long-term frequency stability better than $10^{-12}$.

\end{abstract}

\maketitle

\section{\label{sec:level1}Introduction}

Miniaturized microwave quantum frequency standards based on coherent population trapping (CPT), also known as miniature atomic clocks (MACs) \cite{Vanier_2005}, have small size ($V \,<\, 100$ cm$^3$) and low power consumption ($P \,<\, 1$ W). These features are crucial for many applications of MACs in science and technology. For instance, MACs are in high demand for enhancing the reliability of global navigation satellite systems (GNSS) \cite{Meng_2024}. In conjunction with nanosatellites (CubeSats), MACs enable the exploration of navigation principles for deep space \cite{Nydam_2017} and probing the Earth's ionosphere \cite{Aheieva_2017}. A CPT-based MAC is commonly built using the microwave (``clock'') transition in $^{87}$Rb ($6.8$~GHz) or $^{133}$Cs ($9.2$~GHz) atom, induced by two frequency components of the emission spectrum of a vertical-cavity surface-emitting laser (VCSEL). It means that any resonant microwave radiation is not employed to excite the clock transition (in contrast to double-resonance clocks \cite{Batori_2022}), making this all-optical technology exceptionally compact and energy efficient.

The relative frequency stability of MACs is typically characterized by Allan deviation ($\sigma_y$) \cite{Riehle_2004}. Advanced MACs demonstrate short-term stability at the level of $\sigma_y$$\,\approx\,$$10^{-11}$ at $1$~s averaging and $\sigma_y$$\,\approx\,$$2$$\,\times\,$$10^{-12}$ at $24$~h \cite{Zhang_2019, Skvortsov_2020}. Long-term stability degradation in MACs is generally attributed to the frequency drift of the clock transition, involving the Zeeman sub-states $|F_{g1}, m_{g1} = 0\rangle$ and $|F_{g2}, m_{g2} = 0\rangle$ (the so-called ``0--0'' transition). Here, $F_{g1}$ and $F_{g2}$ stand for the total angular momenta of two hyperfine-splitting energy levels in the atomic ground state (specifically, $F_{g1}$$\,=\,$$1$ and $F_{g2}$$\,=\,$$2$ for $^{87}$Rb). Ones of the major factors contributing to this drift is connected with variations in the optical power ($P$) in the vapor cell and the microwave power ($P_\mu$) of the signal from the microwave synthesizer used to modulate the VCSEL's pump current for obtaining the required spectrum of radiation for CPT resonance excitation. The influence of both factors on the clock transition frequency has a common physical nature and is often termed as light shift \cite{Levi_2000,Miletic_2012}. Other major contributors to the clock's frequency drift are variations in the temperature ($T$) of the atoms and variations in the magnetic field (${\bf B}$) in the vapor cell.

\begin{figure*}
    \centering
    \includegraphics[width=1\linewidth]{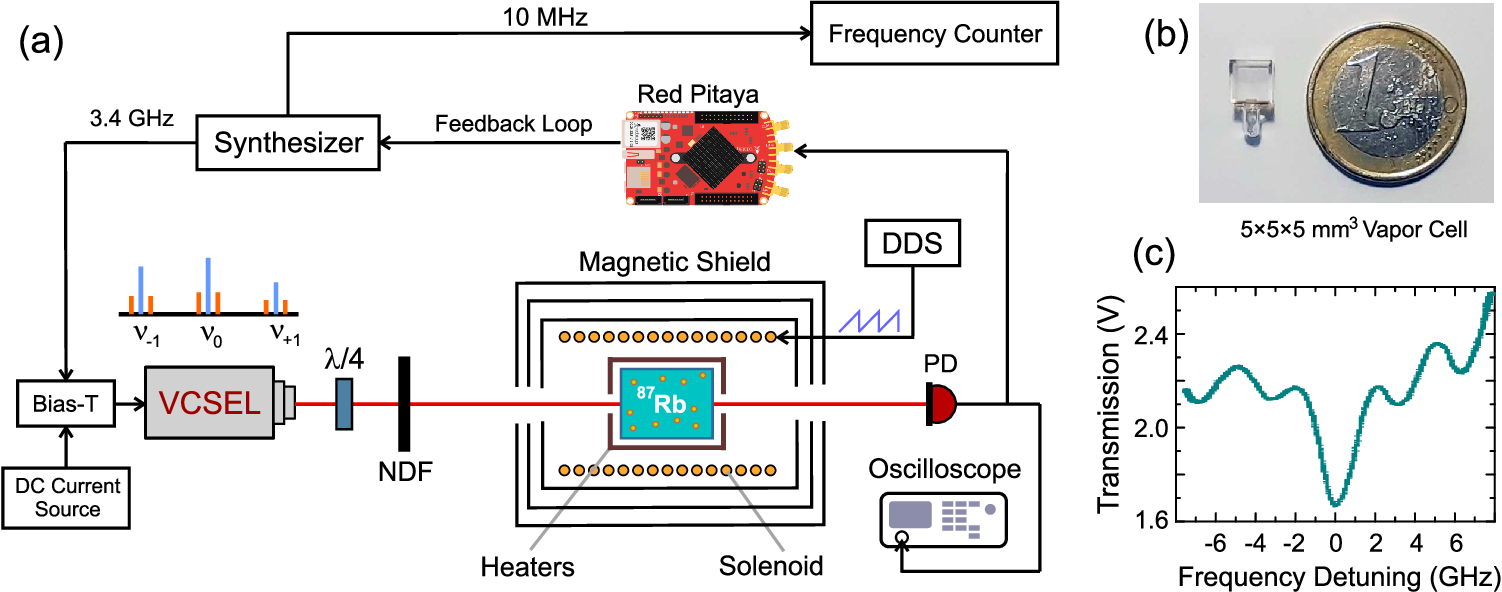}
    \caption{(a) Experimental setup: VCSEL -- vertical-cavity surface-emitting laser, $\lambda/4$ -- quarter-wave plate, NDF -- neutral density filter, PD -- photodetector, DDS -- direct digital synthesizer. (b) Photograph of a rubidium vapor cell. (c) Transmission signal at PD as the function of optical frequency detuning (the microwave $3.4$~GHz modulation of the laser pump current is turned on).}
    \label{fig:1}
\end{figure*}

Thus, to further improve performance of MACs, it is required to develop novel approaches that would reduce the sensitivity of the clock transition frequency to variations in key parameters: $T$, $P$, $P_\mu$, and  $\mathbf{B}$. Beyond the obvious but insufficient approach based on stabilizing each of these parameters individually, a unified strategy exists for more effectively suppressing the influence of these variations on frequency stability of MACs. For instance, when using a particular buffer gas mixture, the dependence of the frequency shift ($\Delta$) of the clock transition on the temperature ($T$) of atoms exhibits an extremum at a specific value $T$$\,=\,$$T_0$ \cite{Vanier_1982,Kozlova_2011,Masian_2015,Skvortsov_2020}, which determines the operating temperature of the vapor cell. Near this extremum, the function $\Delta(T)$ becomes quadratic, making the clock frequency significantly less sensitive to small temperature variations compared to any other value $T$$\neq$$T_0$.

Similarly, one can find a specific microwave power $P_\mu$$\,=\,$$P_{\mu0}$ at which the function $\Delta(P_\mu)$ exhibits an extremum, making the clock transition insensitive to small perturbations in $P_\mu$ \cite{Levi_2000,Vaskovskaya_2019,Skvortsov_2020}. Finally, for a certain $P_\mu = P_{\mu_{1}}$, the function $\Delta(P)$ becomes a horizontal line over a wide range of $P$ \cite{Levi_2000,Miletic_2012,Vaskovskaya_2024,Brazhnikov_2024}, implying no sensitivity of the clock transition to variations in $P$. Also, there is a MAC scheme where instead of a straight line, the function $\Delta(P)$ exhibits an extremum \cite{Vishnyakov_2024}.

To mitigate the influence of magnetic field variations on the stability of CPT-based atomic clock, a bias magnetic field  ${\bf B}_{bias}$ is employed (typically in the range of $100$$\,-\,$$600$~mG, \cite{Skvortsov_2020,Knappe_2007}), directed along the wave vector ${\bf k}$. We assume that ${\bf k}$ is in turn directed along the quantization axis $z$, so that the bias magnetic field can be denoted as $B_z$. This configuration assists selective excitation of the ``0--0'' transition without affecting other microwave (two-photon) transitions in the atomic ground state. The feature of the ``0--0'' transition consists in its quadratic dependence on the magnetic field in contrast to other two-photon transitions in the ground state. It means that the ``0--0'' transition frequency could be insensitive to small variations in the magnetic field in the vapor cell. However, since a relatively large bias field is applied to the atoms, the clock transition frequency has a linear-like sensitivity to small variations in $B_z$, so that the quadratic Zeeman (QZ) shift reads \cite{Knappe_2007}: $\Delta_{QZ}$$\,=\,$$\varkappa$$(B_z$$+$$\delta B_z$$)^2$$\,\approx\,$$\varkappa$$B_z^2$$+$$2$$\varkappa$$B_z$$\delta B_z$, where $\delta B_z$ is the small perturbation in $B_z$, i.e., $\delta B_z$$\,\ll\,$$B_z$ (here, $\varkappa$$\,\approx\,$$575$~Hz/G for $^{87}$Rb \cite{Steck_2024}). Variations in other Cartesian components ($\delta B_x$, $\delta B_y$) still have a small quadratic-manner influence on the clock transition frequency.

For a long time, the magnetic-field-induced frequency shift in MACs was addressed solely through passive magnetic shielding of the vapor cell, achieving a suppression factor at best of $\approx\,$$1000$ \cite{Knappe_2007}. At the same time, in the case of miniaturized (chip-scale) atomic clocks, this becomes a challenging technical task. The situation is further complicated by the presence of stray magnetic fields from various electronic components of the MAC itself, particularly from the vapor cell heaters \cite{Knappe_2007}.  Therefore, in the development of next-generation MACs with long-term frequency stability better than $10^{-12}$, the suppression of MFS is not merely a technical problem but it requires novel approaches. 

In Ref. \onlinecite{Tsygankov_2021}, the authors revealed that the function $\Delta(B_z)$ has an extremum as well, which can be used as an optimal value of the bias magnetic field to reduce influence of variations in ambient magnetic field on the clock transition frequency. The location of the extremum, however, could not be easily adjusted. It was just a single extremum, existing only for a certain type of the circular polarization of the light beam (left-handed or right-handed, depending on the direction of $B_z$). 

In the present work, to lock the microwave frequency of the synthesizer to the clock transition in $^{87}$Rb, we employ a technique, which is analogues to the Pound-Drever-Hall (PDH) technique widely used in the optical frequency range \cite{Black_2001}. According to this technique, the modulating frequency $f_m$ significantly exceeds the resonance linewidth (in MACs, it is typically of the order of $1$~kHz). The PDH-like technique has been previously applied to study CPT resonances in several works \cite{Affolderbach_2000, Kitching_2000,Ben-Aroya_2007,Mikhailov_2010,Yudin_2017,Kobtsev_2018,Kobtsev_2019,Pati_2021,Yudin_2023}. Therefore, here, we do not discuss its advantages. Instead, we show that the PDH technique provides additional capabilities for significant suppression of MFS in MACs.

First, the PDH technique provides additional extrema in the function $\Delta(B_z)$. These extrema exist for both left-handed and right-handed circular polarizations regardless of the bias magnetic field direction. Therefore, there is a choice of the best extremum for the use in MAC. Second, the locations of the extrema can be easily controlled by changing the frequency $f_m$. The latter is of a principal importance for optimizing the atomic clock's regime of operation. Third, the smoothness of the extrema can be high enough to provide immunity of the clock transition frequency to small variations in the magnetic field. The results obtained can be used for developing CPT-based atomic clocks with improved long-term frequency stability.

\begin{figure}[!t]
    \centering
    \includegraphics[width=0.9\linewidth]{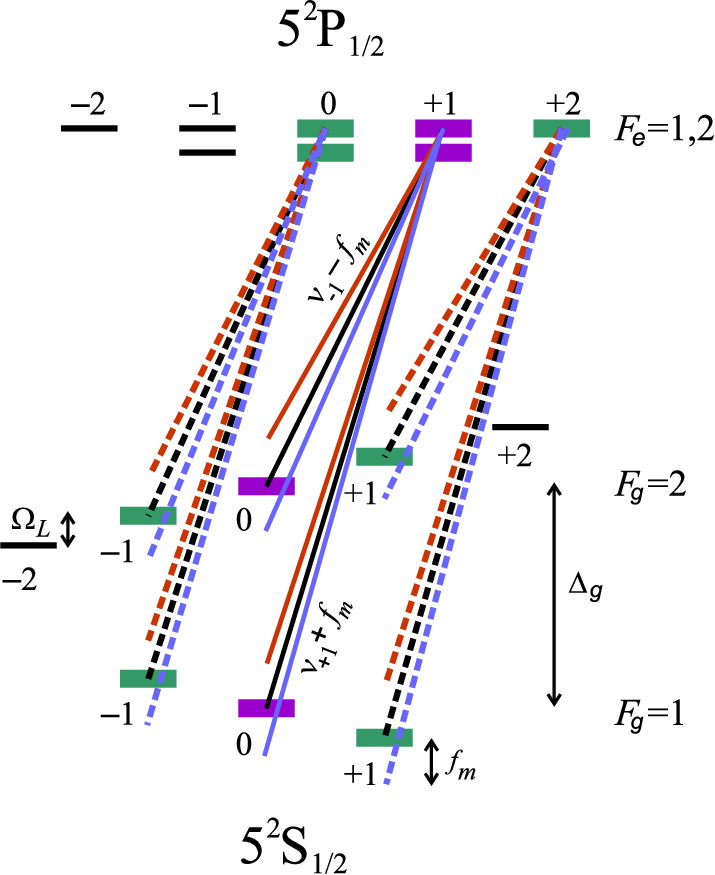}
    \caption{Schematic energy level diagram of the D$_1$ line in $^{87}$Rb. Numbers at Zeeman sub-levels denote magnetic quantum numbers $m$. The Zeeman sub-levels involved in the formation of CPT resonances are highlighted in bold: purple sub-levels for the clock (``0--0'') CPT resonance, while green sub-levels belong to other $\Lambda$-schemes, which are responsible for observing the CPT resonances with high magnetic-field sensitivity. Vertical lines represent $\sigma^+$ transitions induced by the light field: solid lines depict transitions involved in forming the clock CPT resonance, while dashed lines show those contributing to magnetically sensitive CPT resonances. Sidebands with frequencies $v_n$$-$$f_m$ are shown in red, and those with frequencies $v_n$$+$$f_m$ are shown in blue ($n$$\,=\,$$\pm1$).
}
    \label{fig:2}
\end{figure}

\section{Experiments}
\subsection{Experimental setup} \label{sec:level2}

The experimental setup is shown in Fig. \ref{fig:1}(a). A commercially available vertical-cavity surface-emitting laser (\mbox{VCSEL}) with a wavelength of $\approx\,$$794.8$~nm (the D$_1$ line of $^{87}$Rb), a spectrum width of $\approx\,$$50$~MHz, and a maximum optical power of $\approx\,$$250$~$\mu$W was employed as the light source. The laser beam diameter ($e^{-2}$) in our experiments was approximately $1.5$~mm.  In all measurements, the optical power at the cell entrance was maintained at $\approx\,$$17$~$\mu$W using a neutral density filter (NDF). Laser wavelength tuning could be achieved by either adjusting the dc current supply or by controlling the temperature using a Peltier element. The laser was driven by a "Bias-T" scheme, which allowed mixing the dc pump current with a signal at microwave frequency of $\approx\,$$3.417$~GHz from the frequency synthesizer. As the result of this, the VCSEL output radiation was frequency-modulated (FM) and contained a carrier frequency ($\nu_0$) and sidebands of various orders ($\nu_n$, where $n = \pm 1, \pm 2, \dots$). The emission spectrum is schematically depicted in Fig. \ref{fig:1}(a) as a blue-colored comb above the laser.

The experiments were carried out with the use of a cubic $0.125$~cm$^3$ vapor cell made of borosilicate glass [Fig. \ref{fig:1}(b)], filled with $^{87}$Rb atoms and argon as a buffer gas ($\approx\,$$100$ Torr). The cell temperature was maintained at $\approx\,$$330$~K with an accuracy of about $5$~mK using a stabilization electronics, which included film resistive heaters consisted of bifilar copper traces on a polyimide substrate with a $150$~kHz operating frequency. Temperature of the cell was measured using an NTC sensor. The employed thermostabilization electronics had no noticeable impact on the CPT resonance shift and was previously used for high-sensitivity atomic magnetometry \cite{Makarov_2025}.

A three-layer magnetic shield made of permalloy was employed to isolate the cell from external magnetic fields. The residual magnetic field at the center of the shield was on the order of $1$~mG. Orthogonal components ($B_x$, $B_y$) of this field were compensated with the help of Helmholtz coils installed inside the shield. To generate a homogeneous bias magnetic field ($B_z$) in the cell parallel to the light beam propagation, we used a solenoid installed inside the shield. This field split the ground-state hyperfine levels in the atom (the Zeeman effect), enabling selective excitation of the ``0--0'' clock transition (Fig. \ref{fig:2}). The splitting (Larmor) frequency in the ground state of $^{87}$Rb (in Hz) is given by  $\Omega_L$$\,=\,$$\gamma$$B_z$, where $\gamma$$\,\approx\,$$700$~Hz/mG is the gyromagnetic ratio.

The laser frequency was locked to the center of the absorption profile [Fig. \ref{fig:1}(c)] in the same vapor cell where the CPT resonance was observed. A standard synchronous modulation-demodulation method was used for this purpose where the laser frequency was slowly ($\sim\,$$0.5$~Hz) modulated by changing the temperature of the laser diode.

\subsection{\label{sec:level2.2}Measurements and discussions} 

To excite the CPT resonances, we employ the $\nu_{-1}$ and $\nu_{+1}$ optical sidebands of the laser emission spectrum, while other frequencies $\nu_n$ ($n$$\,\neq\,$$\pm 1$) are detuned far enough from the resonance with the medium. At zero Raman (two-photon) frequency detuning, $\delta_R$$\,=\,$$\nu_{+1}$$-$$\nu_{-1}$$-$$\Delta_g$$\,=\,$$0$, atoms are pumped into CPT state and a "dark" resonance is observed in the intensity of the light passed through the vapor cell. Here, $\Delta_g$$\,\approx\,$$6.834$~GHz is the hyperfine splitting frequency of the atomic ground state, and $\nu_{+1}$$-$$\nu_{-1}$$\,=\,$$2f_\mu$ with $f_\mu$ being the microwave modulation frequency of the VCSEL pump current. In particular, solid vertical lines connecting violet Zeeman sub-levels in Fig. \ref{fig:2} correspond to the case when $\delta_R$$\,=\,$$0$.

\begin{figure}[!t]
    \centering
    \includegraphics[width=1\linewidth]{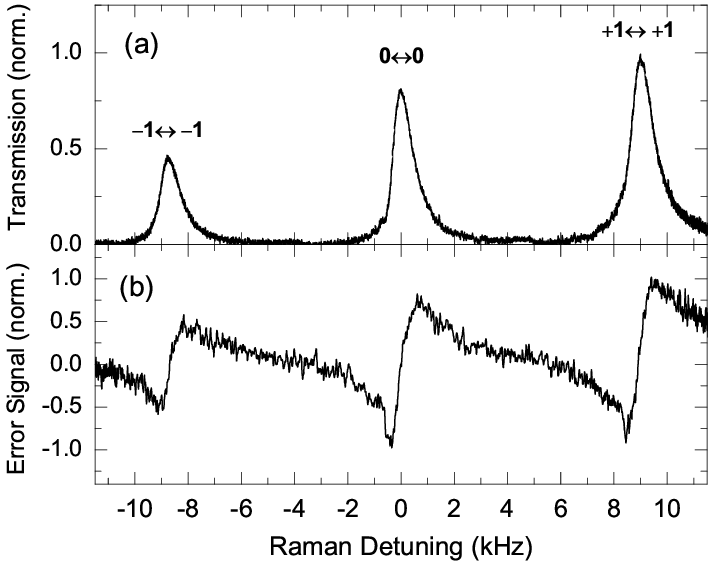}
    \caption{(a) CPT Resonances in a laser beam with right-handed circular polarization. The resonance at center corresponds to the clock transition ``0--0'', left and right resonances are linked to the magnetically sensitive two-photon transitions in the ground state (see Fig. \ref{fig:2}). (b) An error signal formed from the CPT resonances by the lock-in amplifier at $f_m$$\,=\,$$100$~Hz. Here, $B_z$$\,\approx\,$$6.4$~mG. The linewidth of the central resonance is $\approx\,$$900$~Hz.}
    \label{fig:3}
\end{figure}

While real atomic energy levels consist of many Zeeman sub-levels, a simplified three-level ($\Lambda$) spectroscopic model is often employed to qualitatively describe the CPT phenomenon and related effects \cite{Arimondo_1996}. For instance, in the case of the $^{87}$Rb D$_1$ line, when atoms are excited by a bichromatic laser beam, for example, with right-handed circular polarization, the optical transition scheme can be represented as a set of three $\Lambda$ schemes (Fig. \ref{fig:2}). In the presence of a longitudinal magnetic field $B_z$, each scheme contributes to the formation of a CPT resonance at a certain Raman detuning, as demonstrated in Fig. \ref{fig:3}(a). The different amplitudes of CPT resonances are explained by the differences in the Rabi frequencies for the respective $\sigma^+$ transitions in the $\Lambda$ schemes, which are in turn proportional to Clebsch-Gordan coefficients. Note that Fig. \ref{fig:2} displays Zeeman energy level splitting solely in the ground state, as the hyperfine structure is not spectroscopically resolved in the excited state due to collisional broadening. Indeed, as seen from Fig. \ref{fig:1}(c), the absorption linewidth in the vapor cell was approximately $2$~GHz, consistent with data from Ref. \onlinecite{Pitz_2014}, whereas the excited-state hyperfine-splitting frequency is about $0.815$~GHz \cite{Steck_2024}.

\begin{figure}[!t]
    \centering
    \includegraphics[width=1\linewidth]{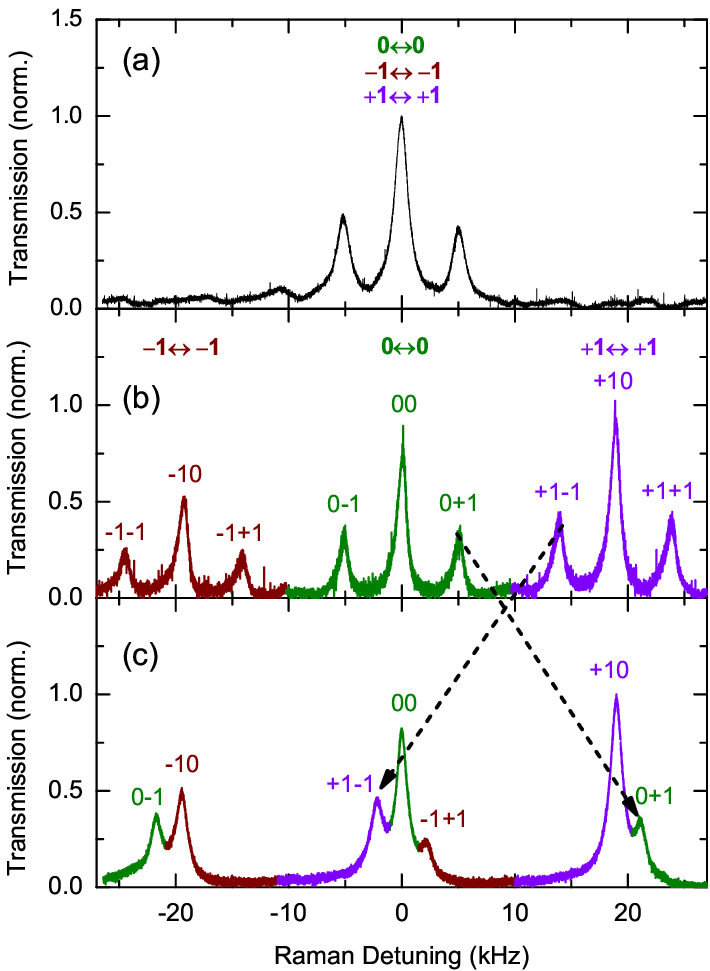}
    \caption{(a) A triple CPT resonance in the PDH regime of excitation at $f_m$$\,=\,$$5$~kHz. The bias magnetic field is absent. All three two-photon transitions in the atomic ground state are simultaneously excited. (b), (c) Two series of CPT resonances in the PDH regime at $B_z$$\,\approx\,$$13.4$~mG when $f_m$$\,=\,$$5$~kHz (b) and $f_m$$\,=\,$$21$~kHz (c).}
    \label{fig:4}
\end{figure}

A conventional synchronous modulation-demodulation method with low modulation frequency ($f_m$$\,=\,$$100$~Hz) for frequency stabilization provides an error signal shown in Fig. \ref{fig:3}(b). The central dispersion-like signal can be used to lock the microwave frequency to the ``0--0'' transition frequency. If the PDH stabilization technique is employed, each optical frequency sideband $\nu_n$ acquires additional low-frequency sidebands at $\nu_n$$\pm$$f_m$. These sidebands are schematically shown in Fig. \ref{fig:1}(a) as red-colored frequency components in the laser radiation spectrum. As shown in Fig. \ref{fig:4}a, these additional frequencies result in observation of a triple CPT resonance corresponding to the ``0--0'' transition (see also plots in Refs. \onlinecite{Vishnyakov_2024,Tsygankov_2024,Pati_2021}). Such a signal can be treated as a sum of three Lorentzian curves. Switching on the bias magnetic field leads to observation of three sets of CPT resonances as shown in Fig. \ref{fig:4}b. There are two sets on the left and on the right side of the plot colored in red and violet, respectively. These sets of resonances originate from left and right CPT resonances in Fig. \ref{fig:3}(a) and are responsible for magnetically sensitive (``non-clock'') two-photon transitions in the atomic ground state: $|F_g$$=1$$,$$\,m_g$$=$$-1$$\rangle$$\,\to\,$$|F_g$$=$$2,$$\,m_g$$=$$-1$$\rangle$ and $|F_g$$=$$1,$$\,m_g$$=$$1$$\rangle$$\,\to\,$$|F_g$$=$$2$$,$$\,m_g$$=$$1$$\rangle$. A central (green) set of CPT resonances is caused by the magnetically insensitive ``0--0'' transition. The resonances in Fig. \ref{fig:4}(b) have been obtained in the regime when $f_m$$\,<\,2\Omega_L$. The regime with $f_m$$\,\geq\,2\Omega_L$ is shown in Fig. \ref{fig:4}(c), when different CPT resonances influence each other. As we will see, the latter has a principle importance for suppressing the MFS.

Table \ref{tab:1} helps to understand which spectral components of the driving laser field are responsible for each CPT resonance shown in Fig. \ref{fig:4}(b) upon changing the Raman frequency detuning. For instance, ``-1-1'' resonance is induced by two spectrum components $\nu_{-1}$$-$$f_{m}$ and $\nu_{+1}$, as well as by other pair of components $\nu_{-1}$ and $\nu_{+1}$$+$$f_{m}$. Neighboring ``-10'' resonance can be induced by three pairs of frequency components: $\nu_{-1}$$-$$f_{m}$ and $\nu_{+1}$$-$$f_{m}$, $\nu_{-1}$ and $\nu_{+1}$, $\nu_{-1}$$+$$f_{m}$ and $\nu_{+1}$$+$$f_{m}$. Let us note that the PDH technique gives higher-order FM sidebands, such as $\nu_{+1}$$\pm$$2f_m$, $\nu_{-1}$$\pm$$2f_m$, etc. Some of them can be distinguished as negligible peaks in Fig. \ref{fig:4}(a). The amplitudes of these components are relatively small under the experimental conditions used, therefore, we do not consider them further.

\begin{table}[!t]
%\centering
\caption{Components of the VCSEL emission spectrum contributing to the formation of CPT resonances shown in Fig. \ref{fig:4}b upon scanning the Raman frequency detuning.}\label{tab:1}
\begin{ruledtabular}
\begin{tabular}{cc}
 Resonance Notation & Responsible Sidebands    \\
 \hline
 $-1-1$ & ($\nu_{-1}$$-$$f_{m}$, $\nu_{+1}$) \\ 
 $+1-1$ & ($\nu_{-1}$, $\nu_{+1}$$+$$f_{m}$) \\ \hline
 
 \,\,\,\,\,$00$  & ($\nu_{-1}$$-$$f_{m}$, $\nu_{+1}$$-$$f_{m}$) \\
 $-10$ & ($\nu_{-1}$, $\nu_{+1}$) \\
 $+10$ & ($\nu_{-1}$$+$$f_{m}$, $\nu_{+1}$$+$$f_{m}$) \\ \hline
        
 $-1+1$ & ($\nu_{-1}$$+$$f_{m}$, $\nu_{+1}$) \\ 
 \,\,\,\,\,$0+1$  & ($\nu_{-1}$, $\nu_{+1}$$-$$f_{m}$) \\ \hline
 
 \,\,\,\,\,$0-1$   & ($\nu_{-1}$$-$$f_{m}$, $\nu_{+1}$)\\ 
 $+1-1$  & ($\nu_{-1}$, $\nu_{+1}$$+$$f_{m}$) \\

\end{tabular}
\end{ruledtabular}
\end{table}

A microwave frequency stabilization feedback loop was established using the ``Red Pitaya'' tool. The CPT resonance frequency shifts were measured at different $f_m$. For each modulation frequency, the phase of the reference signal during demodulation and the frequency deviation were adjusted to maximize the error signal slope and maintain optimal short-term frequency stability of the clock.

\begin{figure}[!b]
    \centering
    \includegraphics[width=1\linewidth]{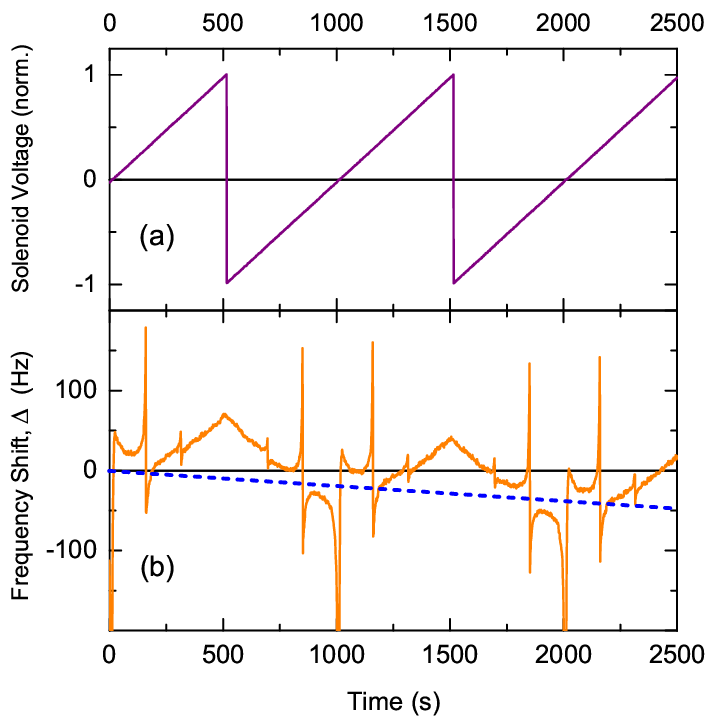}
    \caption{(a) Sawtooth waveform voltage applied to the solenoid to generate a magnetic field $B_z$ in the vapor cell. (b) Frequency shift of the clock transition in the PDH technique upon varying the solenoid voltage. $f_m$$\,=\,$$75$~kHz. Dashed line denotes slow frequency drift.}
    \label{fig:5}
\end{figure}

To study the clock CPT resonance shift versus the magnetic field $B_z$, a variable voltage with a sawtooth waveform was applied to the solenoid [Fig. \ref{fig:5}(a)]. The synthesizer microwave frequency ($\approx\,$$6.834/2$$\,=\,$$3.417$~GHz) was locked to the central ``0--0'' resonance referred to as ``00'' in Fig. \ref{fig:4}(b). To measure the resonance shift, we used a low-frequency output of the synthesizer, which frequency ($10$~MHz) was tightly linked to the microwave frequency via a fixed coefficient. Applying the phase comparator-analyzer ``VCH-323'', this low-frequency signal was compared to a reference signal of the hydrogen rf standard ``CH1-1007'' (``Vremya-CH'' JS Com.). Both devices are denoted as ``Frequency Counter'' in Fig. \ref{fig:1}(a). The magnetically-induced frequency shift of the ``0--0'' transition is shown in Fig. \ref{fig:5}(b). There are several spikes separated by a smooth dependence.

\begin{figure}[!t]
    \centering
    \includegraphics[width=1\linewidth]{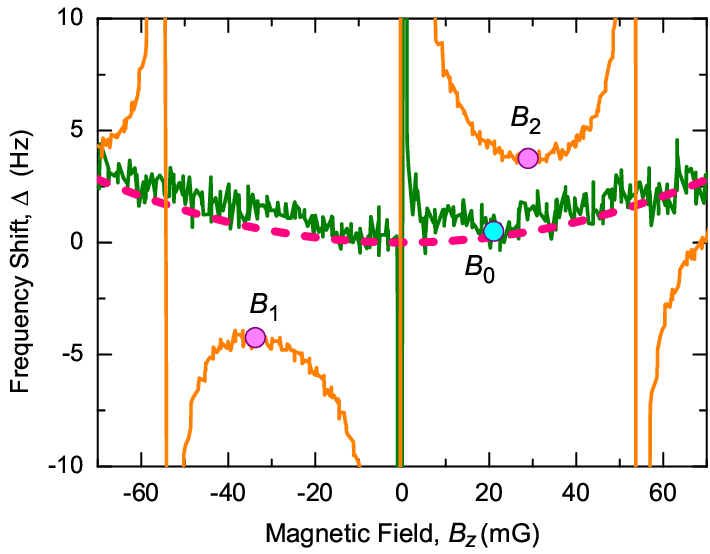}
    \caption{Frequency shift of the ``0--0'' clock transition in the PDH technique (orange solid curve) as the function of the magnetic field $B_z$ at $f_m$$\,=\,$$75$~kHz. Green curve shows the result in a conventional low-frequency modulation technique ($f_m$$\,=\,$$200$~Hz). Dashed pink curve reflects the quadratic Zeeman effect according to Eq. (\ref{shift1}).}
    \label{fig:6}
\end{figure}

Fig. \ref{fig:6} shows a zoomed part of Fig. \ref{fig:5}(b), while the $x$-axis of the plot reflects real magnetic field values. As seen from Fig. \ref{fig:6} (orange solid curve), the function $\Delta(B_z)$ exhibits several extrema, some of which can be used for suppressing the clock transition sensitivity to the external magnetic field variations. Here, we study properties of two extrema denoted as $B_1$ and $B_2$. They attract an attention because they are smooth and are located at relatively small magnetic fields. If there were no spikes, then the function $\Delta(B_z)$ would behave in accordance with the quadratic Zeeman (QZ) effect \cite{Vanier_2005,Steck_2024}:

\begin{equation}\label{shift1}
    \Delta_{\rm QZ}(B_z) \approx 575 \, B_z^2,
\end{equation}

\noindent where $B$ is expressed in G, while the shift is expressed in Hz. This quadratic law is shown in Fig. \ref{fig:6} as dashed red curve.

Sharp peculiarities (spikes) in Fig. \ref{fig:6} located at $B_z$$\,\approx\,$$0$ and $B_z$$\,\approx\,$$\pm$$53$~mG originate mainly from frequency pulling effect \cite{Tsygankov_2021}, implying that the location of the clock ``00'' resonance in Figs. \ref{fig:4}(b) and (c) is influenced by the Lorentzian wings of neighboring resonances during the magnetic field scan. For instance, in Fig. \ref{fig:6}, a spike located at $B_z$$\,\approx\,$$53$~mG is associated with the frequency pulling of the clock $00$ resonance by Lorentzian wings of magnetically sensitive resonances ``+1-1'' and ``-1+1'' as seen in Fig. \ref{fig:4}(c). For comparison, Fig. \ref{fig:6} also reflects the behavior of $\Delta(B_z)$ when a conventional low-modulation-frequency regime of stabilization is applied (green solid curve). As seen from the figure, the function $\Delta(B_z)$ deviates from the quadratic Zeeman law only in vicinity of $B_z$$\,\approx\,$$0$, exhibiting just a single extremum at $B_z$$\,=\,$$B_0$$\,\approx\,$$20$~mG. The similar behavior was observed in Ref. \onlinecite{Tsygankov_2021}.

There are two key advantages of the PDH technique against the conventional one. First, the PDH technique provides a few extrema instead of a single extremum ($B_0$). These extrema appear for both positive and negative magnetic field values, i.e. there is a choice between them. It is noteworthy that the resonances presented in Figs. \ref{fig:3} and \ref{fig:4} will not change when both the sign of the magnetic field ($B_z$$\,\to\,$$-B_z$) and the light polarization sign ($\sigma^+$$\,\to\,$$\sigma^-$) will be simultaneously reversed. This indicates that the same results can be obtained for the left-handed circularly polarized light. This findings distinguishes the PDH technique from the case of low-modulation frequency regime, where the function $\Delta(B_z)$ does not exhibit any extrema for left-polarized light at $B_z$$\,>\,$$0$ (see Ref. \onlinecite{Tsygankov_2021}).

\begin{figure}[!t]
    \centering
    \includegraphics[width=1\linewidth]{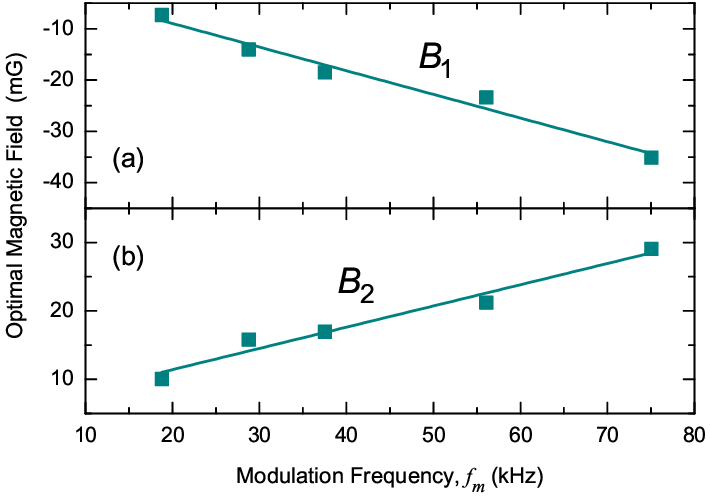}
    \caption{Locations of the extrema  ${\rm B}_1$ (a) and ${\rm B}_2$ (b) versus the modulation frequency $f_m$.}
    \label{fig:7}
\end{figure}

The second advantage of PDH technique is even more relevant for application in MACs. This technique provides an efficient way for controlling the extrema locations. This capability is attractive for real-world applications in MACs as it allows for optimizing their performance. In our experiments, the locations of the extrema $B_{1,2}$ were controlled using the modulation frequency $f_m$. As seen from Fig. \ref{fig:7}, the locations of $B_1$ and $B_2$ exhibit an almost linear behavior when varying $f_m$, meaning that their control is quite simple.

\begin{figure}[!b]
    \centering
    \includegraphics[width=1\linewidth]{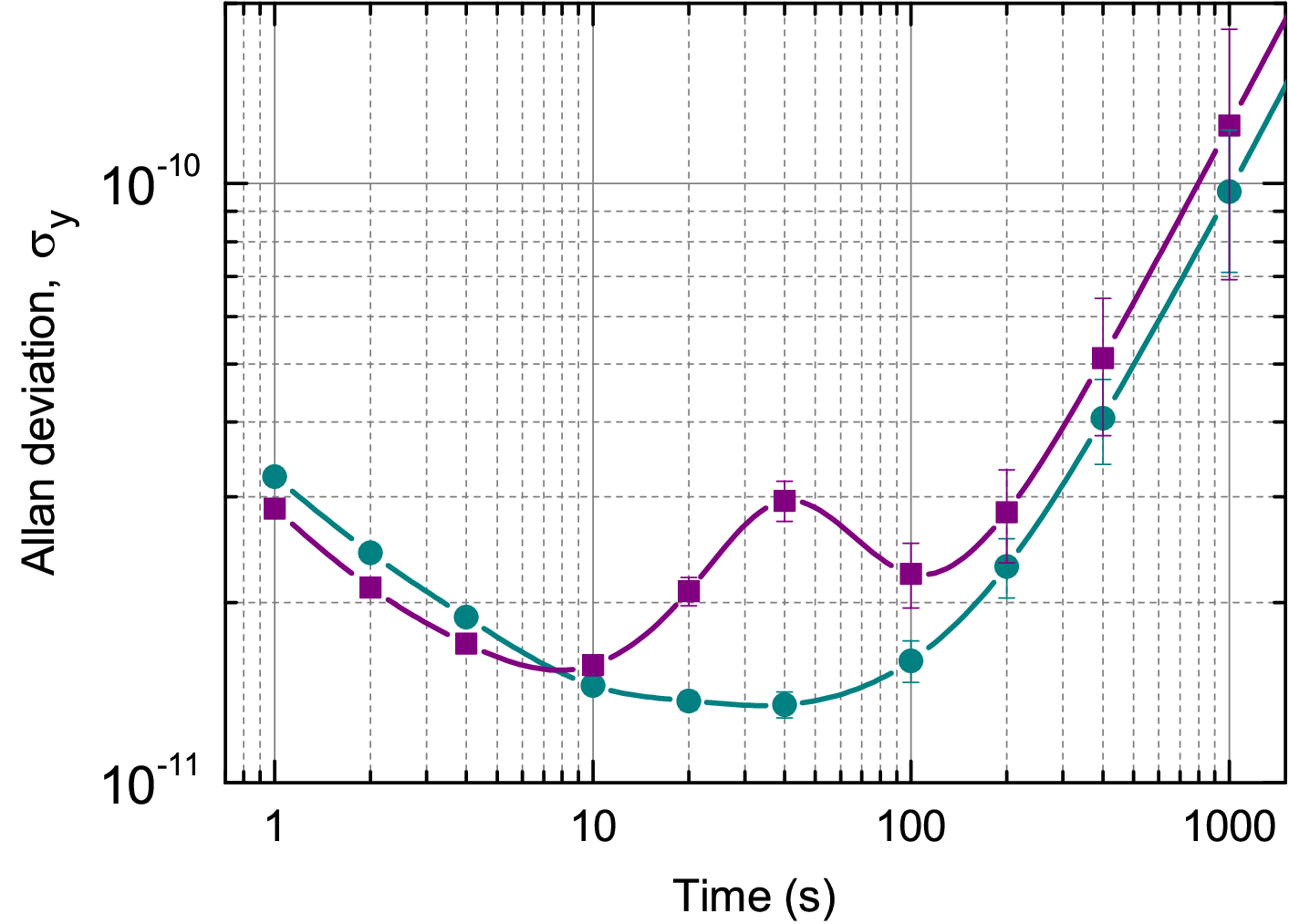}
    \caption{Relative frequency stability of the atomic clock when a $150$~s harmonic magnetic field modulation is applied to the vapor cell. $f_m$$\,=\,$$75$~kHz. The bias field is kept at $B_z$$\,=\,$$B_2$$\,\approx\,$$53$~mG for circles and at $B_z$$\,\approx\,$$100$~mG for squares.}
    \label{fig:8}
\end{figure}

To demonstrate the feasibility of the proposed approach, we have measured the mid-term Allan deviation (Fig. \ref{fig:8}). In the first measurement, the magnetic field was set to the value corresponding to the extremum $B_2$ (blue circles). In the second case, the magnetic field was about $100$~mG where mainly the quadratic Zeeman shift occurs (purple squares). To reveal the influence of magnetic field variations on the clock transition frequency, the magnetic field was harmonically modulated with a period of $150$~s and a peak-to-peak amplitude of $\approx\,2$~mG. Such variations in $B_z$ may occur in a miniature atomic clock, for example, due to changes in their orientation in the Earth's magnetic field ($\approx\,$$500$~mG), if a magnetic shielding factor is $\approx\,$$500$. As seen from Fig. \ref{fig:8}, at $B_z$$\,=\,$$B_2$ (blue circles), the magnetic field variations do not noticeably affect the stability of the clock frequency, which is about $1.3$$\,\times\,$$10^{-11}$ at $50$~s. However, for an arbitrary choice of $B_z$ (purple squares), these oscillations make a visible effect on the stability, degrading it up to $3$$\times$$10^{-11}$.
 
In our experiments, we used a rubidium vapor cell with non-optimized mixture of buffer gases \cite{Vanier_1982,Skvortsov_2020}. Therefore, temperature variations of the atoms could affect the ``0--0'' transition frequency stability at mid and long terms as seen in Fig. \ref{fig:8} (e.g., see Ref. \onlinecite{Liu_2007}). This influence is also seen in Fig. \ref{fig:5}(b) as dashed line. Therefore, to study the mitigation of magnetic-field-induced frequency shift at longer times, a vapor cell with proper buffer gas mixture is required. The workability of the method, nevertheless, is obvious from the obtained results.   

Finally, we can estimate the relative frequency shift of the clock transition resulted from perturbations of the external magnetic field. In the vicinity of values $B_1$ and $B_2$, the function $\Delta(B_z)$ behaves as a parabola. Let us consider, for instance, the extremum $B_2$. An elementary analysis shows that the curvature of the function $\Delta(B_z)$ near $B_z$$\,=\,$$B_2$, i.e., the second derivative $\Delta^{(2)}(B_2)$, determines the relative frequency shift of the clock transition as follows:

\begin{equation}\label{relshift}
    \frac{\delta\Delta}{\Delta_g} \approx \frac{\Delta^{(2)}}{2\Delta_g}\,\delta B_z^2\,,
\end{equation}

\noindent with $\delta B_z$ being a small perturbation of the field $B_z$. From experimental data, using a parabolic approximation in Fig. \ref{fig:6}, we obtain $\Delta^{(2)}(B_2)$$\,\approx\,$$4.4$$\times$$10^{-3}$~Hz/mG$^2$. From (\ref{relshift}), we arrive at the following estimate: $\delta\Delta$$/$$\Delta_g$$\,\approx\,$$3.2$$\times$$10^{-13}$$\delta B_z^2$~mG$^{-2}$, where $\delta B_z$ is expressed in mG. This, in particular, means that when using a magnetic shield with a suppression factor of $500$ (or $\approx\,$$1000$ as in Ref. \onlinecite{Knappe_2007}), the influence of the Earth's magnetic field on the relative frequency stability of MAC will be approximately $3.2$$\,\times\,$$10$$^{-13}$ ($\approx\,$$0.8$$\,\times\,$$10$$^{-13}$).

\section{\label{sec:level3}Conclusion}

Magnetically induced frequency shift of the clock ``0--0'' transition in $^{87}$Rb atom was studied. We have shown that applying the Pound-Drever-Hall-like technique for microwave frequency stabilization offers new capabilities for suppressing the influence of magnetic field perturbations in the vapor cell on the ``0--0'' transition frequency and ultimately the frequency stability of CPT-based atomic clock. This technique provides a few optimal values of the bias magnetic field in the cell, at which magnetic field perturbations have a negligibly small influence on the clock's stability. Furthermore, these values can be easily adjusted.

It has been shown that the residual frequency instability can be as low as $\delta\Delta$$/$$\Delta_g$$\,\approx\,$$3.2$$\times$$10^{-13}$$\delta B_z^2$~mG$^{-2}$ with $\delta B_z$ being a small perturbation (in mG). This result means that application of the proposed approach for building a new-generation CPT-based atomic clock will assist achieving long-term frequency stability well below $10^{-12}$, retaining the requirements for small size, weight, and power consumption (SWaP).

\begin{acknowledgments}
This work was supported by the Ministry of Science and Higher Education of the Russian Federation in the frame of state assignment of Institute of Laser Physics SB RAS (Project reg. no.: 121041300256-1, code: FWGU-2021-0001). The theoretical analysis of the experimental results was supported by the Russian Science Foundation (Grant No. 23-12-00195). 
\end{acknowledgments}

\section*{Data Availability Statement}

The data that support the findings of this study are available from the corresponding author upon reasonable request.

\bibliography{article}

\end{document}